\documentclass[prl, preprintnumbers ,showpacs,amsmath,amssymb, superscriptaddress,twocolumn]{revtex4-1}
\usepackage{bm}
\usepackage{amsmath}
\usepackage{amssymb}
\usepackage{amsthm}
\usepackage{amsfonts}
\usepackage{enumerate}
\usepackage{latexsym}
\usepackage{times}
\usepackage{ulem}
\usepackage{graphicx}
\usepackage{color}
\usepackage{makeidx}
\usepackage{ifpdf}

\begin{document}

\title{Tuning the electronic properties of $J_{\rm eff}=1/2$ correlated semimetal in epitaxial perovskite SrIrO$_{3}$}

\author{Jian~Liu} \email{jian.liu@berkeley.edu}
\author{J.-H.~Chu}
\affiliation{Department of Physics, University of California, Berkeley, California 94720, USA}
\affiliation{Materials Science Division, Lawrence Berkeley National Laboratory, Berkeley, California 94720, USA}
\author{C.~Rayan~Serrao}
\author{D.~Yi}
\affiliation{Department of Materials Science and Engineering, University of California, Berkeley, California 94720, USA}
\author{J.~Koralek}
\affiliation{Materials Science Division, Lawrence Berkeley National Laboratory, Berkeley, California 94720, USA}
\author{C.~Nelson}
\affiliation{Department of Materials Science and Engineering, University of California, Berkeley, California 94720, USA}
\author{C. Frontera}
\affiliation{Institut de Ci\`{e}ncia de Materials de Barcelona (ICMAB-CSIC), Campus de la UAB, E-08193 Bellaterra, Spain}
\author{D. Kriegner}
\affiliation{Institute of Semiconductor and Solid State Physics, University Linz Altenbergerstr. 69, A-4040 Linz, Austria}
\author{L. Horak}
\affiliation{Faculty of Mathematics and Physics, Charles University, Ke Karlovu 5, 121 16 Prague, Czech Republic}
\author{E.~Arenholz}
\affiliation{Advanced Light Source, Lawrence Berkeley National Laboratory, Berkeley, California 94720, USA}
\author{J. Orenstein}
\affiliation{Department of Physics, University of California, Berkeley, California 94720, USA}
\affiliation{Materials Science Division, Lawrence Berkeley National Laboratory, Berkeley, California 94720, USA}
\author{A.~Vishwanath}
\affiliation{Department of Physics, University of California, Berkeley, California 94720, USA}
\affiliation{Materials Science Division, Lawrence Berkeley National Laboratory, Berkeley, California 94720, USA}
\author{X.~Marti}
\affiliation{Department of Materials Science and Engineering, University of California, Berkeley, California 94720, USA}
\author{R. Ramesh}
\affiliation{Department of Physics, University of California, Berkeley, California 94720, USA}
\affiliation{Materials Science Division, Lawrence Berkeley National Laboratory, Berkeley, California 94720, USA}
\affiliation{Department of Materials Science and Engineering, University of California, Berkeley, California 94720, USA}

\date{\today}

\begin{abstract}
  We investigated the electronic properties of epitaxially stabilized perovskite SrIrO$_{3}$ and demonstrated the effective strain-control on its electronic structure. Comprehensive transport measurements showed that the strong spin-orbit coupling renders a novel semimetallic phase for the $J_{\rm eff}=1/2$ electrons rather than an ordinary correlated metal, elucidating the nontrivial mechanism underlying the dimensionality-controlled metal-insulator transition in iridates. The electron-hole symmetry of this correlated semimetal was found to exhibit drastic variation when subject to bi-axial strain. Under compressive strain, substantial electron-hole asymmetry is observed in contrast to the tensile side, where the electron and hole effective masses are comparable, illustrating the susceptivity of the $J_{\rm eff}=1/2$ to structural distortion. Tensile strain also shrinks the Fermi surface, indicative of an increasing degree of correlation which is consistent with optical measurements. These results pave a pathway to investigate and manipulate the electronic states in spin-orbit-coupled correlated oxides, and lay the foundation for constructing 5d transition metal heterostructures.
\end{abstract}

\maketitle

Strongly spin-orbit-coupled quantum materials have received enormous attention recently since the spin-orbit coupling (SOC) is an essential interaction in spintronics and affords a critical pathway to nontrivial symmetry-protected topological states \cite{Hasan2010}. Electronically, this is achieved by removing orbital degeneracy via spin-orbit entanglement and significantly modifying the band structure, such as in topological insulators. On the other hand, electron-electron interaction is fundamental for generating fascinating quantum phenomena in strongly correlated systems, including superconducting cuprates and colossal magnetoresistive manganites. A critical ingredient therein is also lifting orbital degeneracy of correlated electron but through coupling of the $d$-states to the lattice distortion, e.g. the crystal field effect. Intriguing situations arise when both SOC and correlation are of similar strength in a single system. To this end, many theoretical proposals have been put forward for obtaining a topological Mott insulator, Weyl semimetal, quantum anomalous Hall effect, spin liquid, unconventional superconductivity, and so on both in the bulk and heterostructures \cite{Pesin2010,Xiao2011,Ran2009,Wan2011,Wang2011}.

\begin{figure}[b]\vspace{-0pt}
\includegraphics[width=8.5cm]{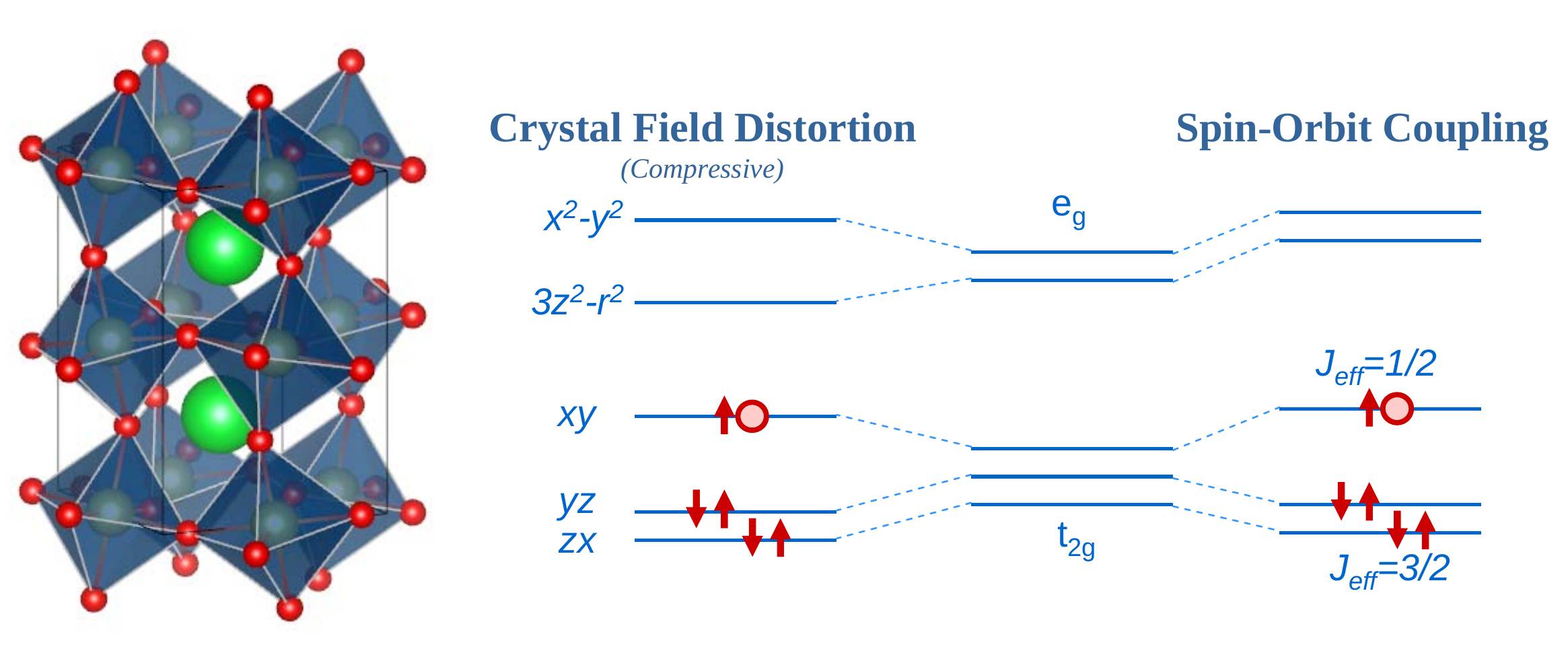}
\caption{\label{schematic} (color online) Left: orthorhombic unit cell of SrIrO$_3$. Right: schematic of $d$-orbital energy levels under strong SOC or (compressive) crystal field distortion.}
\end{figure}

The fundamental challenge for realizing these fascinating properties, however, is to understand the role of SOC across the Mott metal-insulator transition. 5$d$ transition metal oxides, particularly the iridates, are prototypes under these circumstances. Under an octahedral crystal field, the five $d$-states are split into the $e_g$ and $t_{2g}$ manifolds. While a tetragonal distortion may further remove the orbital degeneracy, the strong SOC could also split the three-fold degenerate $t_{2g}$ orbitals into a $J_{\rm eff}=1/2$ doublet and a $J_{\rm eff}=3/2$ quadruplet, where the spin is coherently integrated (see Fig.~\ref{schematic}). The Ir$^{4+} $$5d^5$ low-spin configuration, thus, leads to a half-filled $J_{\rm eff}=1/2$ band which undergoes a metal-insulator transition in the Sr$_{n+1}$Ir$_{n}$O$_{3n+1}$ ($n$ is integer) family with decreasing dimensionality \cite{Moon2008}. The $n=1$ endmember Sr$_{2}$IrO$_{4}$ has received enormous attention due to its antiferromagnetic insulating state \cite{Kim2008,Kim2009,Kim2012} that may be significantly modified by octahedral distortion \cite{Jackeli2009,Haskel2012,Serrao2013}, while the $n=\infty$ endmember, perovskite SrIrO$_{3}$ (see Fig.~\ref{schematic}), is believed to be an ordinary correlated metal in proximity to a Mott instability \cite{Moon2008}. Recent theoretical calculations, however, suggest that the strong SOC and correlation may results in a $J_{eff}=1/2$ correlated semimetal band on the itinerant side \cite{Zeb2012}, raising the question on the nature of the Mott transition in the strong SOC-limit. Moreover, this unique SOC-induced semimetallic state has been suggested to be a key component for building novel topological insulators in SrIrO$_{3}$-based artificial heterostructures \cite{Xiao2011,Carter2012}. Verification of this picture and achieving control of this electronic state have been lacking because bulk SrIrO$_{3}$ forms the 6H hexagonal structure instead of the perovskite phase (which is stable only in polycrystalline form under high pressure) \cite{Longo1971,Jang2010}.

In this letter, we demonstrate the semimetallic behavior in the heteroepitaxy-stabilized perovskite phase of SrIrO$_{3}$ and the remarkably tunable electron-hole symmetry of the $J_{eff}=1/2$ band. The highly epitaxial perovskite phase of SrIrO$_{3}$ was stabilized within a wide range of lattice mismatch. This strain enables control on the semimetallicity and the SOC- and correlation-derived electronic structure. Semimetallic behavior of the $J_{\rm eff}=1/2$ band, such as small carrier densities, was confirmed by transport measurements. A two-carrier model analysis reveals a drastic variation of the electron-hole mobility symmetry modulated by strain. Assuming an equal scattering rates for electron and hole, the results indicate that the compressive strain favors larger Fermi surfaces where the electrons have a smaller effective mass than the holes, whereas tensile strain tends to maintain small Fermi surfaces and high symmetry of the electron and hole effective masses, signifying latent nontrivial band crossing. Tensile strain is also highly effective in tuning the degree of correlation as shown by optical measurements.

\begin{figure}[t]\vspace{-0pt}
\includegraphics[width=8.5cm]{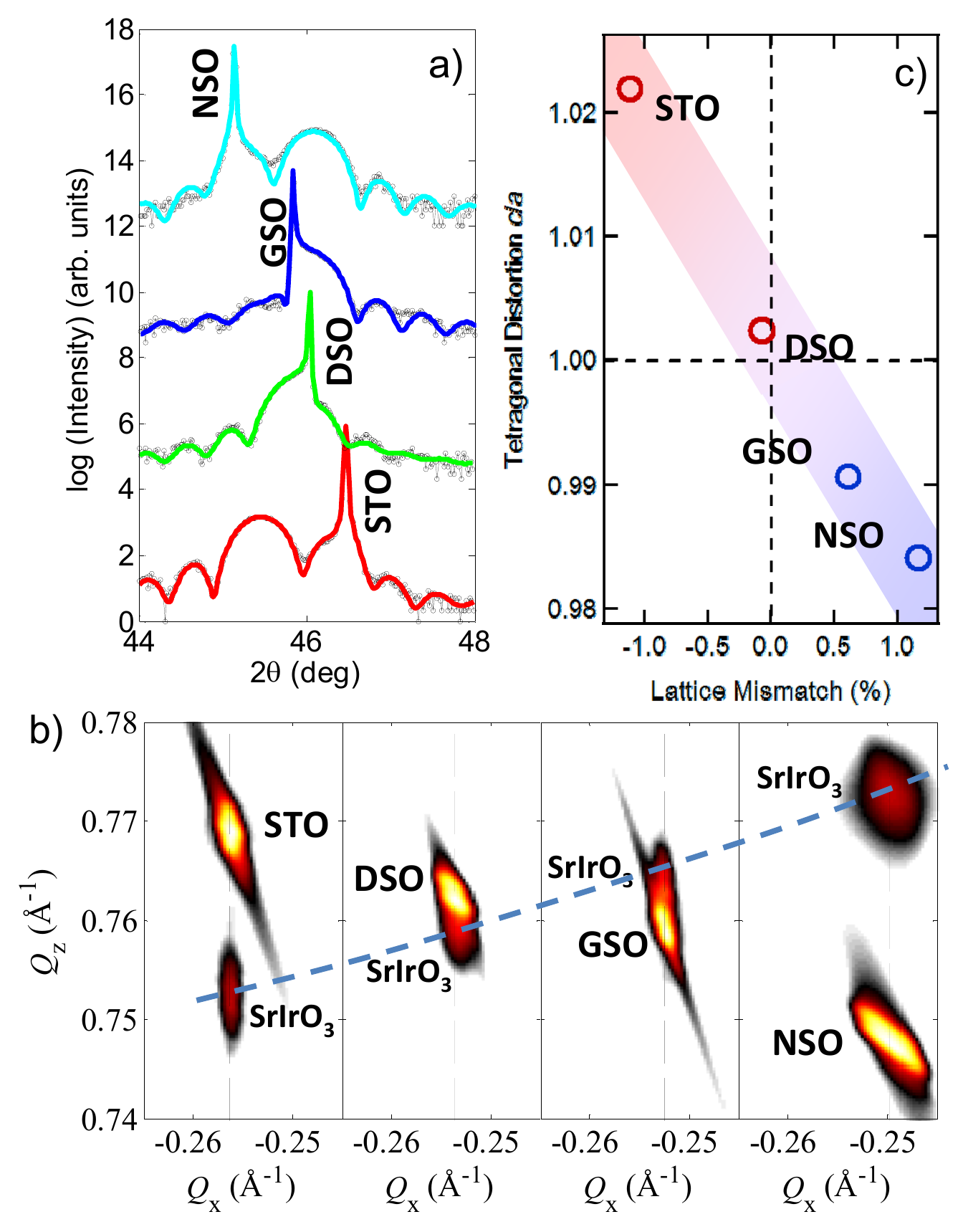}
\caption{\label{XRD} (color online) (a) Specular x-ray diffraction scans and corresponding dynamical theory diffraction simulations around the pseudo-cubic (002) peak. (b) Reciprocal space maps around the pseudo-cubic (103) reflection ((-103) for the sample on NSO). Dashed lines are guides for the eye. (c)structural tetragonal distortion, defined as the ratio between the out-of-plane and in-plane pseudo-cubic lattice parameters $c/a$, versus the lattice mismatch.}
\end{figure}

Here we utilized epitaxial stabilization to obtain high-quality single-crystal thin films of perovskite SrIrO$_{3}$ by pulsed laser deposition.
In addition, a series of pseudo-cubic (001)-oriented perovskite substrates, including SrTiO$_3$ (STO), DyScO$_3$ (DSO), GdScO$_3$ (GSO), and NdScO$_3$ (NSO), was used to apply lattice mismatch up to $1.2$\%. The film thickness in this study was set to 18 to 20 nm confirmed by x-ray reflectivity measurements. Specular scans from high-resolution x-ray diffraction at room temperature indicate all samples are single-phase with all the pseudo-cubic (00$l$) peaks of the perovskite structure. More importantly, the (002) peak shown in Fig.~\ref{XRD}(a) exhibits a systematic shift with the substrate lattice size, indicating a shrinking out-of-plane lattice parameter $c$ accordingly from STO, DSO, GSO, to NSO. The observed total thickness oscillations also corroborates the coherently grown structures. These scans are in accordance with the dynamical theory diffraction simulations consisting of the substrate and an 18nm-thick layer of SrIrO$_{3}$. Reciprocal space mapping around the pseudo-cubic (103) peak (see Fig.~\ref{XRD}(b)) reveals a fully strained state to the substrate in all samples, which is crucial for epitaxial stabilization of the perovskite phase. Based on the extracted lattice parameters, we quantify the amount of strain-induced bi-axial tetragonal distortion by the ratio of $c$ and $a$, with 1 corresponding to the undistorted case. As shown in Fig.~\ref{XRD}(c), the degree of tetragonal distortion displays a clear dependence on the lattice mismatch. In particular, the samples are compressively strained with $c/a$ equal to 1.022 and 1.002 on STO and DSO, respectively, albeit the latter is very close to the undistorted case. On the other hand, the samples on GSO and NSO exhibit tensile distortion with $c/a$ equal to 0.991 and 0.984, respectively.

\begin{figure*}[t]\vspace{-0pt}
\includegraphics[width=5.9cm]{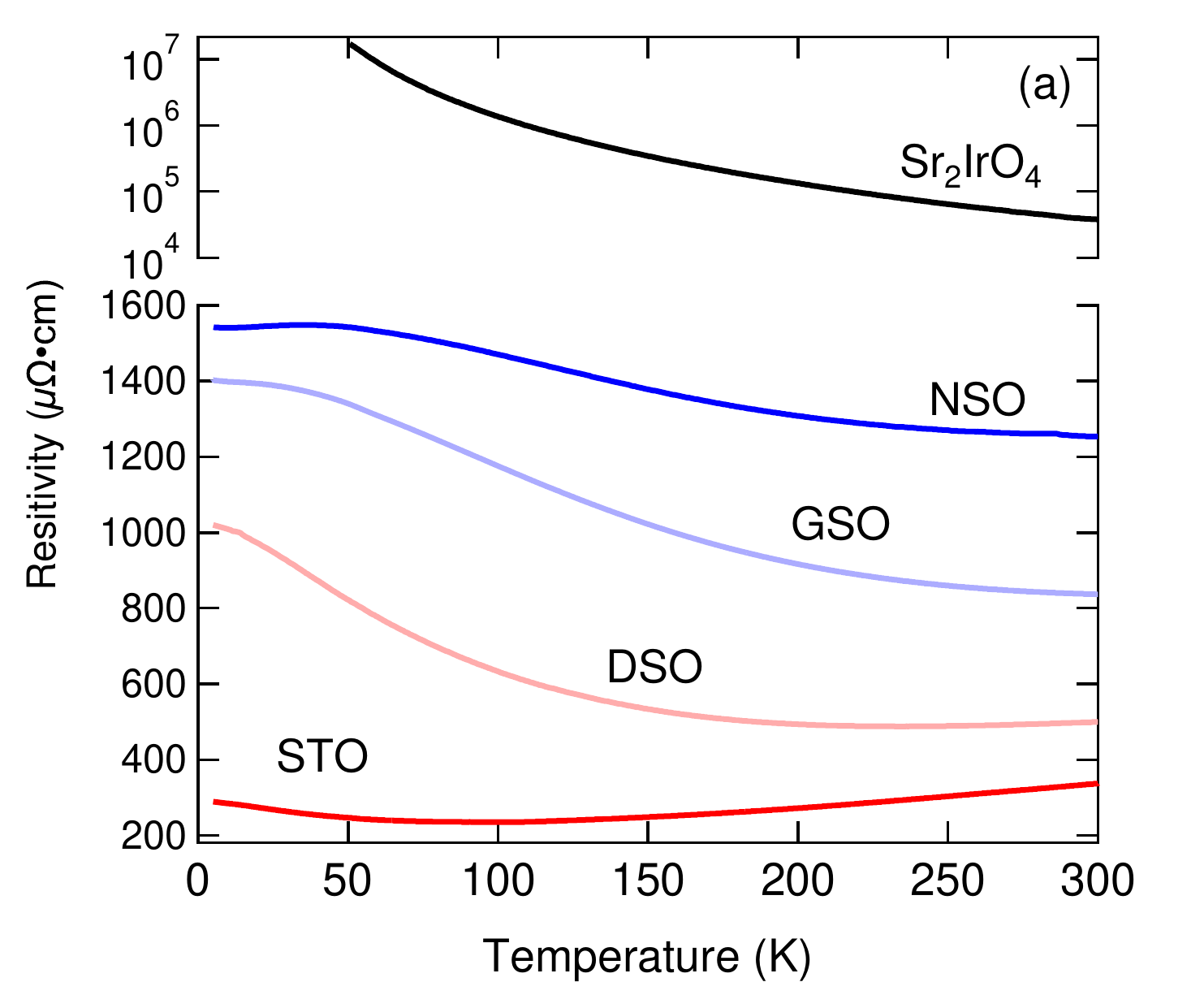}
\includegraphics[width=5.9cm]{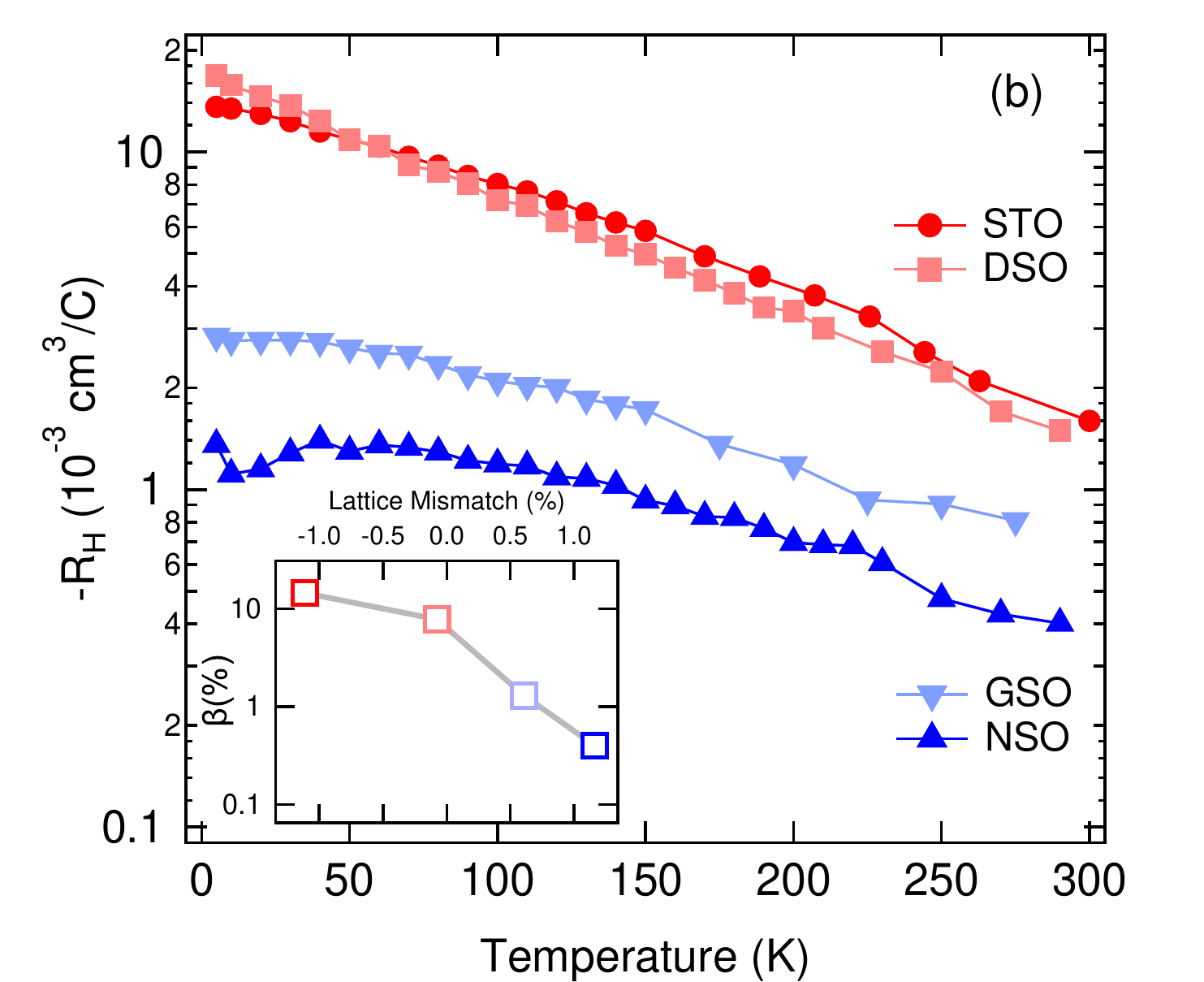}
\includegraphics[width=5.9cm]{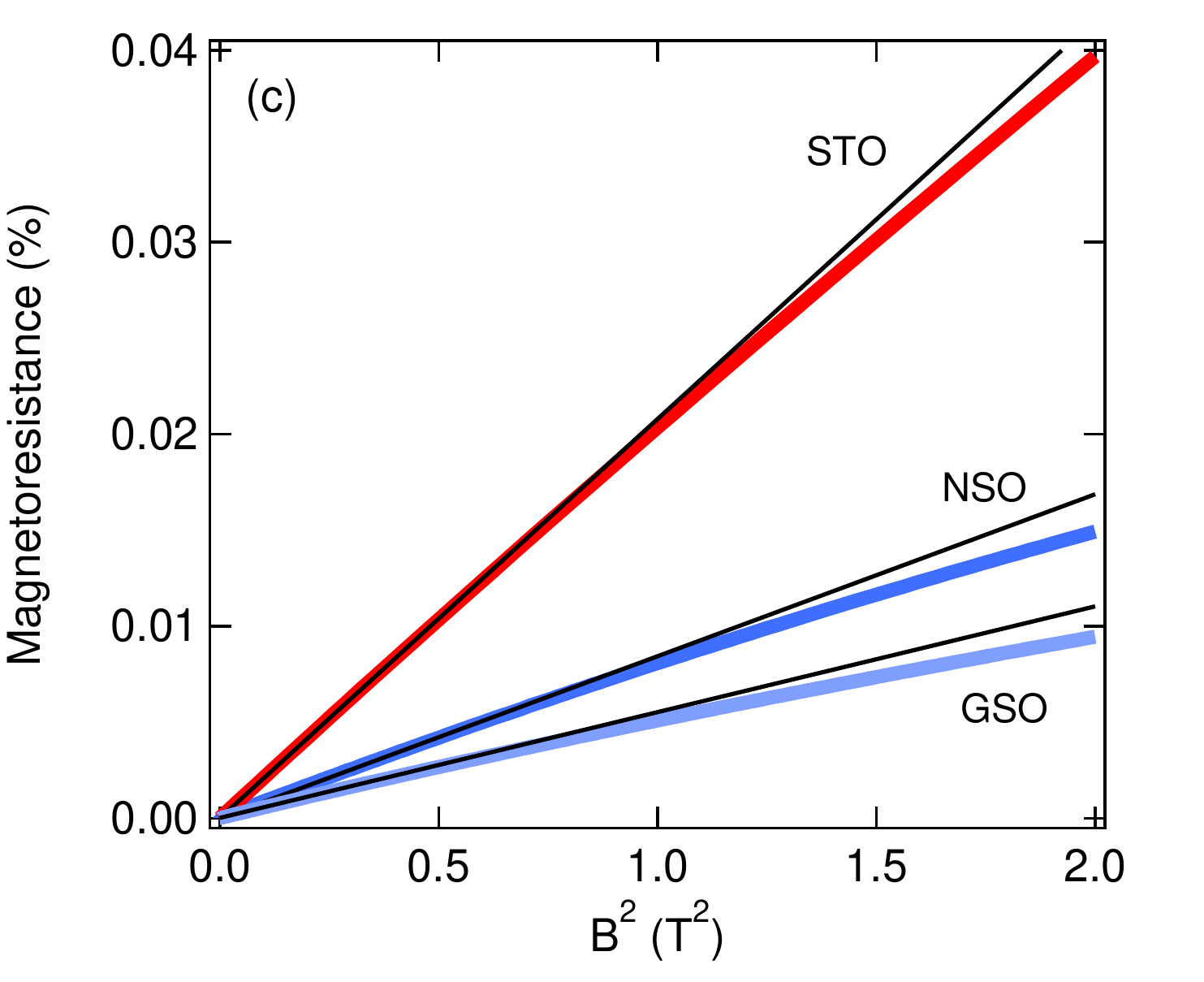}
\caption{\label{transport}
(color online) Resistivity (a) and Hall coefficient (b) as a function of temperature for samples on STO, DSO, GSO, and NSO. Resistivity of a 60 nm Sr$_2$IrO$_4$ is included for comparison. Inset of (b) shows the electron-hole asymmetry versus lattice mismatch at 10 K. (c) Magnetoresistance (thick lines) versus $B^2$ at 10 K on STO, GSO and NSO. Thin black lines are guide for the eye, indicating the linear dependence on $B^2$ in the low field limit.}
\end{figure*}

Figure~\ref{transport}(a) shows the temperature-dependent resistivity $\rho$(T) of the samples on the four substrates from 5 to 300 K, where an evolution is evident as a function of the tetragonal distortion. In particular, the sample on STO shows a clear metallic behavior at high temperatures as $\rho$(T) decreases with reducing temperature, i.e. $d\rho/dT>0$. Below $\sim$100 K, $\rho$(T) reaches a minimum and gradually increases ($d\rho/dT<0$), resulting in a small upturn. Note that, while a similar behavior with a resistivity minimum around 50 K has been reported in bulk polycrystal \cite{Zhao2008}, the overall resistivity of the sample on STO here is one order of magnitude smaller ($\sim300$ ${\rm \mu\Omega\cdot cm}$ v.s. $\sim2-4$ $m{\rm \Omega\cdot cm}$ at 300 K) \cite{Longo1971,Zhao2008}. With reducing compressive strain, a similar, albeit weaker, metallic behavior is observed for the sample on DSO, accompanied by a higher upturn temperature ($\sim$230 K). Although the resistivity increase at low temperatures is also stronger, no divergent behavior is observed; $\rho$(5K) $\approx1$ $m{\rm \Omega\cdot cm}$ is only about twice of $\rho$(300K), suggesting the absence of a true charge excitation gap. This is in sharp contrast to the strong insulating behavior in Sr$_2$IrO$_{4}$ shown in Fig.~\ref{transport}(a) \cite{Serrao2013}.
 On the tensile side where $c/a$ is less than 1, while the upturn temperature is further raised to around 300 K, the sample on GSO also exhibits a slow resistivity increase of a similar magnitude to that on DSO. An interesting difference, however, emerges at low temperatures where the resistivity increase slows down as $|d\rho/dT|$ clearly starts to drop from $\sim$50 K. This behavior sets in around 100 K in the sample on NSO, and a resistivity maximum is reached around 50K followed by a reentrance of a weak metallic behavior.

From the data above one can see that this system can neither be described as an insulator nor a normal metal within the studied range of lattice mismatch. The gradual change of resistivity within one order of magnitude together with the presence of resistivity minimum and maximum can be a signature of strong competition between the carrier density and mobility, characteristic of semimetallic systems. To disentangle these two contributions, we measured the Hall coefficient $R_{\rm H}$ which is shown in Fig.~\ref{transport}(b) as a function of temperature for the samples on the four substrates.  While, in a compensated system such as the half-filled $J_{\rm eff}=1/2$ band in SrIrO$_{3}$, the Hall effect is usually vanishingly small because the electron and hole contributions cancel, the observed $R_{\rm H}$ under compressive strain is negative and relatively large indicative of significant electron-hole asymmetry. In addition, $|R_{\rm H}|$ also display a strong strain-dependence as large as an order of magnitude at low temperatures.

To obtain quantitative insight, the Hall coefficient was analyzed together with the resistivity in a two-carrier model with an equal number of $J_{\rm eff}=1/2$ electrons and holes. $\rho$ and $R_{\rm H}$ in this case are given by
\begin{equation}\label{rho}
    \rho=\frac{1}{2ne\mu}
\end{equation}
\begin{equation}\label{RH}
    R_{\rm H}=-\frac{\beta}{ne}
\end{equation}
with
\begin{align*}
    \beta&=\frac{\mu_{e}-\mu_{h}}{\mu_{e}+\mu_{h}} & \text{and}& & \mu&=\frac{\mu_{e}+\mu_{h}}{2}
\end{align*}
where $n$ is the electron/hole carrier density, $e$ is the electron charge, $\mu$ is the average of the electron mobility $\mu_{e}$ and the hole mobility $\mu_{h}$, and $\beta$ is their asymmetry. The negative sign of $R_{\rm H}$, thus, indicates $\beta>0$, i.e. $\mu_{e}>\mu_{h}$. More importantly, since $|R_{\rm H}|\propto\beta$, the large change in $|R_{\rm H}|$ induced by strain can be attributed to a strongly modulated electron-hole asymmetry, in addition to a changing $n$ which characterizes the size of the Fermi surface. This can be further seen in the ratio of $|R_{\rm H}|$ and $\rho$; $\frac{|R_{\rm H}|}{\rho}=2\beta\mu=\mu_e - \mu_h$ is essentially the mobility difference between electron and hole. Since $|R_{\rm H}|$ drops by one order and $\rho$ becomes as much as four times larger at low temperatures from the compressive to tensile strain, e.g. STO vs NSO, $\mu_e-\mu_h$ must reduce by $\sim$ a factor of 1/50 to take into account this difference. Thus, assuming $\mu$ is relatively unchanged, the electron mobility must be largely enhanced relative to hole mobility from the tensile to the compressive side, suggesting that the $J_{\rm eff}=1/2$ band structure is significantly by the bi-axial tetragonal distortion.

To verify this picture, we measured magnetoresistance ($\frac{\rho(B)-\rho(0)}{\rho(0)}$) with the magnetic field $B$ along the surface normal. In the low-field limit, magnetoresistance is described by the classical quadratic $B^2$-dependence with a coefficient $A$ written as
\begin{equation}\label{A}
    A=(1-\beta^{2})\mu^2
\end{equation}
in the two-carrier model for the half-filled $J_{\rm eff}=1/2$ state. This behavior can be seen in Fig.~\ref{transport}(c) which shows the magnetoresistance at 10 K versus $B^2$. By combining Eqs.~\ref{rho} - \ref{A}, one can disentangle $\beta$ and $\mu$, and also solve for $n$, which are summarized in Table~\ref{table}. In particular, the carrier density $n$ was found to be at the level of $\sim$10$^{19}$ cm$^{-3}$ corresponding to $\sim$0.001 electron/hole per Ir. This carrier density is much smaller than that of typical metals ($\sim$10$^{23}$ cm$^{-3}$), but comparable with semimetals such as graphite, antimony as well as antiferromagnetic iron pnictides. Its variation with strain, e.g. from $\sim$2.3$\times$10$^{19}$ cm$^{-3}$ on NSO to $\sim$6.7$\times$10$^{19}$ cm$^{-3}$ on STO, also confirms the semimetallic state with a small Fermi surface tunable by the bi-axial distortion.
The existence of the semimetallic ground state manifests the critical role of spin-orbit coupling in differentiating the dimensionality-controlled metal-insulator transition of iridates from the conventional Mott transition in transition metal compounds. Namely, the correlated gap is open from semimetallic bands that only have a small density of states around Fermi level but significant spectral weight depleted to finite energies. Note that, semimetallic behavior is also found in the itinerant-to-localized crossover of pyrochlore iridates \cite{Ueda2012}, and can be characteristic of the interplay between strong SOC and correlation in vicinity of a metal-insulator transition.

In addition to the small Fermi surface, the obtained electron-hole asymmetry also displays a remarkably strong strain-dependence (see inset of Fig.~\ref{transport}(b)); for instance, $\beta$ is equal to 0.4\% and 14.5\% on NSO and STO, respectively.
At low temperatures, the dominant scattering comes from disorder, independent of the type of carriers. Since mobility is the inverse of the product of scattering rate and effective mass, the electron-hole asymmetry is due to their different effective masses.
With this condition, the highly modulated electron-hole symmetry of the $J_{\rm eff}=1/2$ states demonstrates that the semimetallic dispersion and bandwidth can be selectively tuned by epitaxial strain-controlled distortion. Specifically, tensile strain favors a highly symmetrical semimetallic band structure with a smaller band overlap. In contrast, electron and hole bands are strongly asymmetrized under compressive strain, indicating electron and hole pockets of drastically different topologies emerge at the Fermi surface. Microscopically, this high sensitivity of the semimetallicity signifies the prominent control of the interplay between the SOC and the bi-axial distortion to the IrO$_6$-octahedral network on the orbital degree of freedom of the $J_{\rm eff}=1/2$ states and their intersite hopping. Note that, on the basis of the present measurements, it is not possible to distinguish a trivial semimetal arising from closing of an indirect gap from a more interesting Dirac-like dispersion or coexistence of both \cite{Zeb2012}, which is a topic of future study. Since both inversion and time reversal symmetry appear to be present, a Weyl semimetal is however not expected \cite{Wan2011}. But, given the results above, a clean Dirac-like dispersion typically characterized by high electron-hole symmetry and small carrier density is more likely to be found under larger tensile strain.

\begin{table}[h]
\caption{\label{table} Strain state, bi-axial distortion, carrier density, electron-hole asymmetry, and average mobility of SrIrO$_3$ on the four substrates.}
\begin{ruledtabular}
\begin{tabular}{ccccccc}
Substrate&Strain\footnote{"+" and "-" correspond to compressive and tensile, respectively.}&$c/a$&$n$(10K)&$\beta$(10K)&$\mu$(10K)\\
&&&(cm$^{-3}$)&(\%)&(cm$^2$/(V$\cdot$s))\\
\hline
STO &-&1.022&6.7$\times$10$^{19}$&14.5&163\\
DSO\footnote{$n$, $\beta$ and $\mu$ on DSO were estimated by assuming a magnetoresistive coefficient $A$ similar to that on STO.} &-&1.002&3.1$\times$10$^{19}$&7.8&102\\
GSO &+&0.991&3$\times$10$^{19}$&1.3&74\\
NSO &+&0.984&2.3$\times$10$^{19}$&0.4&92\\
\end{tabular}
\end{ruledtabular}
\end{table}

With increasing temperature, the difference in $|R_{\rm H}|$ between the compressive and tensile sides becomes slightly reduced, suggesting a relatively weaker strain-dependence of the electron-hole asymmetry.
This is expected since both electrons and holes should be subject to inelastic scattering which becomes a dominant factor in mobility at high temperatures. The corresponding reduction in mobility is, however, not reflected in $\rho(\rm T)$ which shows a predominant $d\rho/dT<0$ behavior in wide temperature ranges (except on STO). There thus must be a stronger increase of the carrier density, which is indeed consistent with the reduction in $|R_{\rm H}|$. This prominent competition between the mobility and the carrier density results in the combination of the $d\rho/dT<0$ and $d\rho/dT>0$ behaviors with a hardly changed $\rho$ over the entire temperature range.

\begin{figure}[t]\vspace{-0pt}
\includegraphics[width=8.5cm]{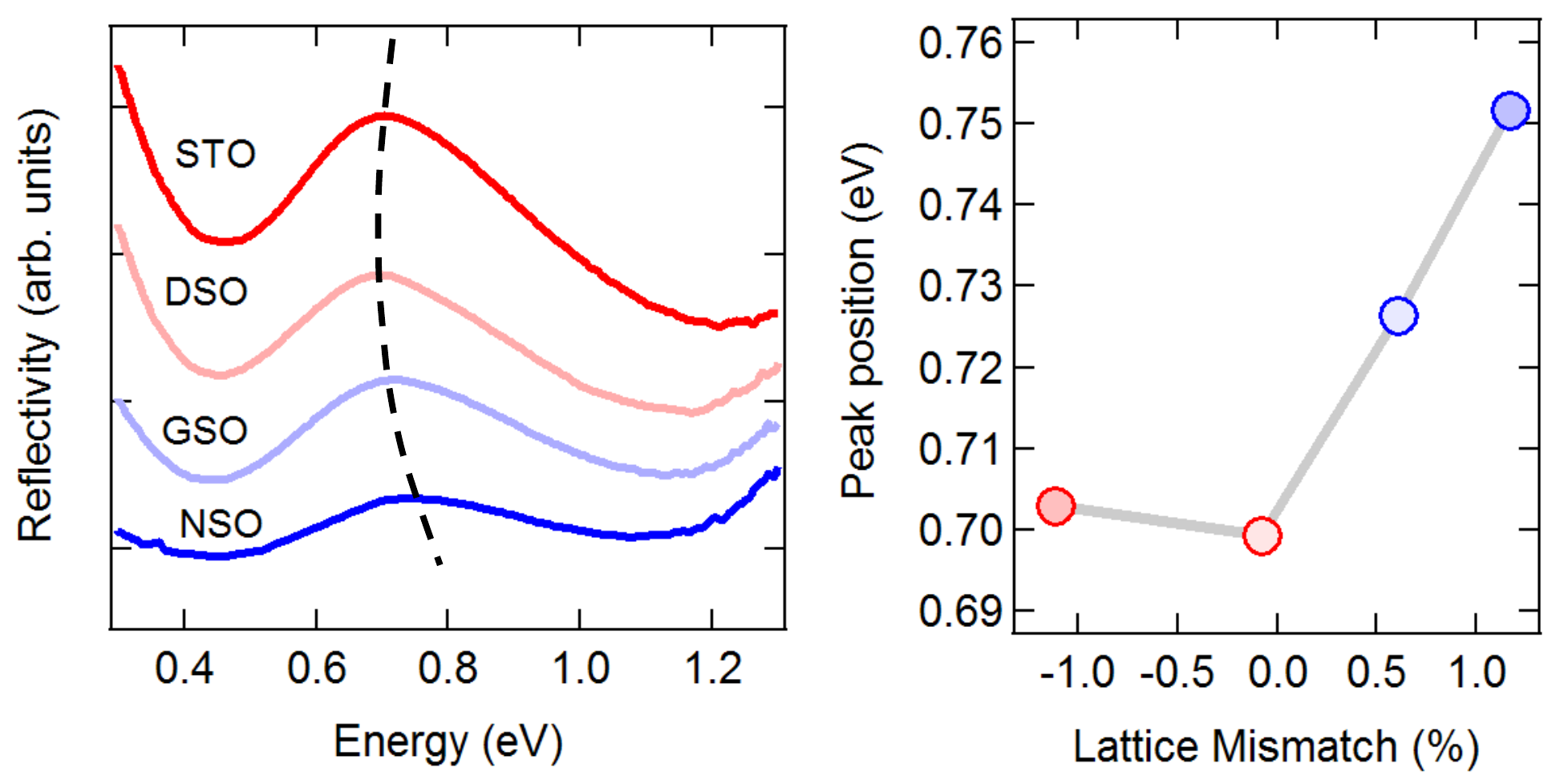}
\caption{\label{FTIR}
(color online) Left: reflectivity spectra from fourier transform infrared spectroscopy at 300 K for samples on STO, DSO, GSO, and NSO. Spectra are shifted vertically for comparison. Dashed curve is guide for the eye. Right: peak position as a function of lattice mismatch.}
\end{figure}

To give further insight into the underlying electronic structure, we performed Fourier transform infrared spectroscopy measurements in the energy range 0.3 - 1.3 eV at 300 K in a Thermo Scientific Nicolet 6700/Continuum XL in the reflection geometry. By virtue of the wide band gaps of the dielectric substrates, the spectral contribution in this region is attributed to SrIrO$_{3}$. From the reflectivity spectra shown in Fig.~\ref{FTIR}, one can clearly see a peak around 0.7 eV corresponding to the interband optical transition from the occupied $J_{\rm eff}=3/2$ states to the empty $J_{\rm eff}=3/2$ states, i.e. the electron band. More importantly, this peak exhibits a clear shift toward higher energies as strain increases on the tensile side; while remaining at a similar position for the samples on STO and DSO, the peak position increases by 25 and 50 $m$eV on GSO and NSO, respectively. This shift indicates a larger separation between the electron and hole band centers under tensile strain, consistent with the smaller band overlap and Fermi surface. Interestingly, this peak energy also increases by 130 $m$eV due to the enlarged Mott gap from the insulating sister compound Sr$_{3}$Ir$_{2}$O$_{7}$ to Sr$_{2}$IrO$_{4}$ \cite{Moon2008}. By the same token, the observed peak shift implies enhanced correlation and strong Mott instability, which is also inline with the reduced carrier density and mobility. It is possible that this many-body effect assisted in maintaining the high electron-hole symmetry, while approaching the Mott transition from the semimetallic phase, by suppressing density of states or localizing pockets originating from conventional band overlapping.

In conclusion, we were able to stabilize single crystal thin films of perovskite SrIrO$_{3}$ and tune its SOC-derived electronic properties in a wide range of epitaxial strain. This provides a pathway to explore the fundamental physics behind the interplay between strong SOC and correlation, and control functionalities not accessible in the bulk. In particular, the $J_{\rm eff}=1/2$ states were demonstrated to form an unusual semimetallic phase in the itinerant side across the Sr$_{n+1}$Ir$_{n}$O$_{3n+1}$ series, highlighting the uniqueness of its metal-insulator transition. When subject to tetragonal distortion to the structure, the electron-hole symmetry of the semimetallic bands exhibits large tunability as much as two orders of magnitude. The electrons become significantly more mobile than the holes and dominate the transport properties under compressive strain. Tensile strain, on the other hand, facilitates a high electron-hole symmetry and reduces the number of carriers by reducing the electron and hole band overlap, indicative of a nontrivial semimetallic state. Meanwhile, the degree of correlation was also found to be increased.
These results build the foundation for tailoring 5$d$ transition metal oxides heterostructures and searching for novel correlated topological quantum states.

The authors thank J. Analytis, P. Ryan and J. W. Kim for insightful discussion. We also thank Michael Martin and Hans Bechtel for assistance with the optical spectroscopy performed at the Advanced Light Source. J.L. thanks the support from the Quantum Material program in the Materials Sciences Division in Lawrence Berkeley National Laboratory. This work is supported by the Director, Office of Science, Office of Basic Energy Sciences, Materials Sciences and Engineering Division, of the U.S. Department of Energy under Contract No. DE-AC02-05CH11231.  We acknowledge additional support of the material synthesis facility through the D.O.D. ARO MURI, E3S, and DARPA.

\end{document}